\documentstyle[aps,preprint,tighten]{revtex}
\begin{document}
\draft
\title{Strong Electron Tunneling through a Small Metallic Grain}
\author{D.S.Golubev$^a$ and A.D.Zaikin$^{a,b}$}
\address{$^a$ I.E.Tamm Department of Theoretical Physics,\\
P.N.Lebedev Physics Institute, Leninskii prospect 53, 117924\\
Moscow, Russia \\
$^b$ Institut f\"ur Theoretische Festk\"orperphysik, Universit\"at\\
Karlsruhe, 76128 Karlsruhe, FRG}
\maketitle

\begin{abstract}
Electron tunneling through mesoscopic metallic grains can be treated
perturbatively only provided the tunnel junction conductances are
sufficiently small. If it is not the case, fluctuations of the grain charge
become strong. As a result (i) contributions of all -- including high energy
-- charge states become important and (ii) excited charge states become
broadened and essentially overlap. At the same time the grain charge remains
discrete and the system conductance $e$-periodically depends on the gate
charge. We develop a nonperturbative approach which accounts for all these
features and calculate the temperature dependent conductance of the system
in the strong tunneling regime at different values of the gate charge.
\end{abstract}


Coulomb effects may have a strong impact on electron transport through small
tunnel junctions and metallic grains \cite{AL,SZ}. Provided the resistance
of tunnel junctions $R_t$ between the grain and the lead electrodes is large
$R_t\gg R_q=\pi /2e^2\simeq 6.5$ K$\Omega $ many features of single electron
tunneling are well described within a simple perturbation theory in a
dimensionless junction conductance $\alpha _t=R_q/R_t$ \cite{AL}. This
theory uses the concept of discrete charge states of the grain and describe
the system dynamics in terms of occupation probabilities of these charge
states.

In the limit $\alpha_t \ll 1$ nontrivial features appear only in the vicinity
of the Coulomb blockade threshold in which case two charge states become nearly
degenerate and the perturbation theory fails. Nonperturbatively this
problem has been treated in Refs. \cite{M,GZ94,SS}, and again the physical
picture of electron tunneling via discrete charge has been exploited.

If the conductance of tunnel junctions is not small $\alpha _t\gtrsim 1$ the
problem turns out to be more complicated. Indeed, already making use of a
simple perturbative formula for the inverse lifetime of the excited grain charge
state $Q=Q_0>e/2$ at $T=0$ (see e.g. \cite{AL,SZ})
 $\Gamma =2\alpha _te(Q_0-e/2)/\pi C$
one can immediately conclude that for $\alpha _t\gtrsim 1$
broadening of the excited charge states $Q\gtrsim e$ due to strong quantum fluctuations
of the charge is of the order of the spacing between them $\Gamma \sim E_C$.
Thus
charge levels overlap and the very concept of tunneling via discrete charge
states with given energies becomes illdefined for such values of $\alpha $.
The finite $T$ effect makes this overlap even more pronounced.

In this Letter we propose a theoretical approach which allows to obtain a
quantitative description of electron transport through mesoscopic metallic
grains in the strong tunneling regime. We reformulate the problem in terms
of the variable canonically conjugated to that of the charge, analyze its
quantum dynamics and obtain an expression for the system conductance valid
for all values of the gate charge and practically all experimentally
relevant values of temperature.

We will consider a standard model for a SET transistor: a small metallic
grain is embedded between two bulk electrodes and connected with them via
tunnel junctions with resistances $R_L$ and $R_R$ and capacitances $C_L$ and
$C_R$ respectively for left and right junctions. A gate voltage $V_g$ is
applied to the grain via a capacitance $C_g$, a transport voltage between
two electrodes is equal to $V_x$. We also assume that the impedance of
the external circuit is much less then quantum resistance. This system can
be described by the Hamiltonian
\begin{equation}
\hat H=(\hat q-Q_g)^2/2C+\hat H_L+\hat H_R+\hat H_g+\hat H_T  \label{ham}
\end{equation}
where $\hat q$ is the operator of the grain charge, $
Q_g=C_LV_L-C_RV_R+C_gV_g $ is the (noninteger) external charge, $
V_{L,R}=V_xR_{L,R}/(R_L+R_R)$ and $C=C_L+C_R+C_g$. The terms $\hat H
_h=\sum_k\epsilon _{kh}a_{kh}^{+}a_{kh}$ describe the kinetic energy of
noninteracting electrons in the left ($h=L$) and right ($h=R$) electrodes
and in the grain ($h=g$), whereas the term
\begin{equation}
\hat H_T=\sum_{h=L,R}\sum_{k,k^{\prime }}T_ha_{kg}^{+}a_{k^{\prime }h}\exp
(-i\hat \varphi _h/2)+c.c.  \label{tunn}
\end{equation}
takes care about electron tunneling between the electrodes and the grain
(the tunneling matrix elements $T_{kk^{\prime }}^{L,R}$ multiplied by the
densities of states yield the junction resistances $R_{L,R}=4\pi
e^2N_{L,R}(0)N_g(0)|T_{L,R}|^2$ ). The junctions phase operators can be
expressed as
\[
\hat \varphi _{L,R}=\frac{2eC_{R,L}}{C_L+C_R}V_xt\mp \frac{2eC_gt}{
C_L+C_R+C_g}\left( \frac{C_R-C_L}{C_R+C_L}\frac{V_x}2-V_g\right) \mp \hat
\varphi ,
\]
where $\hat \varphi $ is the ''phase'' of the grain, or, more exactly, is
the operator canonically conjugated to the grain charge $\hat q$: $\left[
\hat \varphi ,\hat q\right] =2ei.$

After a standard procedure of averaging over the electronic degrees of
freedom (see e.g. \cite{SZ}) one can reformulate the problem in terms of the
reduced density matrix $\rho (\varphi ,\varphi ^{\prime })$ which depends
only on the phase variable $\varphi $. If the charge varies continuously
everywhere in the system the density matrix $\rho _c(\varphi ,\varphi
^{\prime })$ is nonperiodic in $\varphi $ \cite{SZ} and obeys a standard
normalization condition $\int\limits_{-\infty }^{+\infty }d\varphi \rho
_c(\varphi ,\varphi )=1$. In our physical situation, however, the charge on
the grain is quantized in units of the electron charge $e$. In
this case the phase variable is compact (i.e. the states $\varphi $ and $
\varphi +4\pi $ are equivalent) and the density matrix obeys the conditions
\cite{SZ}
\begin{equation}
\rho _d(\varphi _1+4\pi n,\varphi _2+4\pi m)=\exp \left( i\frac{2\pi Q_g}e
(n-m)\right) \rho _d(\varphi _1,\varphi _2),  \label{shift}
\end{equation}
\begin{equation}
\int\limits_{-2\pi }^{2\pi }d\varphi \rho _d(\varphi ,\varphi )=1.
\label{norm2}
\end{equation}

Let us now introduce a nonperiodic in $\varphi $ density matrix $\tilde \rho
(\varphi _1,\varphi _2)$:
\begin{equation}
\rho _d(\varphi _1,\varphi _2)=\sum\limits_{n,m}\exp \left( i\frac{2\pi Q_g}e
(n-m)\right) \tilde \rho (\varphi _1-4\pi n,\varphi _2-4\pi m),  \label{m}
\end{equation}
which satisfies the following normalization condition
\begin{equation}
\sum\limits_n\int\limits_{-\infty }^{+\infty }d\varphi \exp \left( i\frac{
2\pi Q_g}en\right) \tilde \rho (\varphi -4\pi n,\varphi )=1  \label{norm3}
\end{equation}
The matrix $\tilde \rho $ (\ref{m}) obeys the same equation of motion as the
density matrix $\rho _c$ describing the continuous charge distribution in
the system. If we assume that our system is ergodic we can immediately
establish the connection between these density matrices. Indeed, at
sufficiently large time the solution of the linear equation of motion for a
dissipative ergodic system acquire a unique form irrespectively to a
particular choice of the initial conditions. Thus in the long time limit the
two solutions of this equation may differ only by a constant, which can be
fixed with the aid of the normalization condition and we get at $
t\rightarrow \infty $
\begin{equation}
\tilde \rho (t,\varphi _1,\varphi _2)=\frac{\rho _c(t,\varphi _1,\varphi _2)
}{\sum\limits_n\int\limits_{-\infty }^{+\infty }d\varphi \exp \left( i\frac{
2\pi Q_g}en\right) \rho _c(t,\varphi -4\pi n,\varphi )}.  \label{connection}
\end{equation}
The equation (\ref{connection}) can now be used for evaluation of the
expectation value of an arbitrary operator $\hat A(\hat \varphi )$, which is
$4\pi $-periodic in $\varphi $. With the aid of (\ref{m},\ref{connection})
after a simple algebra we obtain
\begin{equation}
\left\langle \hat A\right\rangle _d=\int\limits_{-2\pi }^{+2\pi }d\varphi
A(\varphi )\rho _d(\varphi ,\varphi )=\frac{\sum\limits_n\left\langle \hat A(
\hat \varphi )\exp \left( i\frac{2\pi (Q_g-\hat q)}en\right) \right\rangle _c
}{\sum\limits_m\left\langle \exp \left( i\frac{2\pi (Q_g-\hat q)}en\right)
\right\rangle _c}  \label{av1}
\end{equation}
This is one of the main results of the present paper. It establishes a
straightforward connection between the expectation values for an operator of
any physical quantity calculated for discrete and continuous charge
distributions.

Let us now turn to a calculation of the tunneling current through a SET
transistor. We first obtain a formal expression for the expectation value of
the current operator in our system and then evaluate it with the aid of the
equation (\ref{av1}). The first part of this program will be carried out
within the quasiclassical Langevin Equation approach \cite{S,GZ92} developed
under the assumption that fluctuations of the phase variable are (in some
sense) small. This is a suitable assumption as long as fluctuations of the
charge are large. Expressing the kernel of the evolution operator in terms
of the path integral on a real time Keldysh contour and calculating this
integral within the quasiclassical approximation (see \cite{GZ92} for
further details) we obtain
\begin{equation}
C_{L,R}\frac{\ddot \varphi _{L,R}}{2e}+\frac 1{R_{L,R}}\frac{\dot \varphi
_{L,R}}{2e}-\dot q_{L,R}=\tilde \xi _{R,L}=\xi _{L,R1}(t)\cos \left( \frac{
\varphi _{L,R}}2\right) +\xi _{L,R2}(t)\sin \left( \frac{\varphi _{L,R}}2
\right)  \label{eq}
\end{equation}
where $\dot \varphi _{L,R}/2e$ and $\dot q_{L,R}$ define respectively
fluctuating voltages and currents across the left and the right junctions, $
\xi _{L,R1,2}$ are Gaussian stochastic variables describing the shot noise
in these junctions and obeying the conditions
\begin{eqnarray*}
\left\langle \xi _{R,L1,2}(0)\xi _{R,L1,2}(t)\right\rangle =\frac 1{R_{R,L}}
\int \frac{d\omega }{2\pi }\omega \coth \left( \frac \omega {2T}\right) \exp
\left( i\omega t\right) .
\end{eqnarray*}
As the external impedance is negligible, the phases $\varphi _{L,R}$ are
linked to the transport voltage $V_x$ by means of an obvious equation $\dot
\varphi _L+\dot \varphi _R=2eV_x$. Further relations between the phase and
the charge variables are defined by the charge conservation law $\dot q_L-
\dot q_R=C_g\ddot \varphi _g/2e$ and the Kirghoff's law $\dot \varphi _L/2e-
\dot \varphi _R/2e=2V_g-\dot \varphi _g/e$, $\varphi _g$ is the
gate capacitance phase defined analogously to $\varphi _{L,R}$. Combining all
these equations with (\ref{eq}) after averaging over the stochastic
variables $\xi $ we arrive at the expression for the current in our system:
\begin{equation}
I=\langle \dot q_L\rangle =\frac{V_x}{R_L+R_R}-\frac{R_L\left\langle \tilde
\xi _L\right\rangle _d+R_R\left\langle \tilde \xi _R\right\rangle _d}{R_L+R_R
}  \label{vax1}
\end{equation}

To evaluate the average values in (\ref{vax1}) we shall use the result (\ref
{av1}). Assuming that fluctuations of the charge are Gaussian the
contribution of the $n$-th term to the expectation value (\ref{av1}) can be
roughly estimated as:
\begin{equation}
\left\langle A(\varphi )\exp \left( i\frac{2\pi q}en\right) \right\rangle
_c\sim \exp \left( -\frac{2\pi ^2\left\langle \delta q^2\right\rangle }{e^2}
n^2\right) .  \label{estimation}
\end{equation}
Thus provided the charge fluctuations are not small $\left\langle \delta
q^2\right\rangle \gtrsim e^2$ it is sufficient to leave only the terms with $
n,m=0,\pm 1$ in the expression (\ref{av1}). In this approximation we obtain
\begin{equation}
\ \left\langle \tilde \xi _{R,L}\right\rangle _d=\frac{\ \left\langle \tilde
\xi _{R,L}\right\rangle +2\left\langle \tilde \xi _{R,L}\cos \left( \frac{
2\pi }eq\right) \right\rangle }{1+2\left\langle \cos \left( \frac{2\pi }e
q\right) \right\rangle },  \label{average}
\end{equation}
where $\langle ...\rangle $ denotes the average with the density matrix $
\rho _c$ describing the continuous charge distribution. Making use of the
equations (\ref{vax1}), (\ref{average}) and assuming the phase fluctuations
to be small $|\delta \varphi |\lesssim \pi $ in the limit of small transport
voltages we arrive at the expression for the linear conductance
\begin{equation}
(R_L+R_R)G(T)=1-f(T)-g(T)e^{-F(T)}\cos \left( \frac{2\pi Q_g}e\right),
\label{G}
\end{equation}
where $Q_g=C_gV_g$. We define $\alpha _t=\pi /2e^2R_0$, $1/R_0=1/R_L+1/R_R$ and
\begin{equation}
f(T)=\frac 1{2\alpha _t}\left[ \gamma +\frac{2\alpha _tE_C}{\pi ^2T}\Psi
^{^{\prime }}\left( 1+\frac{2\alpha _tE_C}{\pi ^2T}\right) +\Psi \left( 1+
\frac{2\alpha _tE_C}{\pi ^2T}\right) \right] ,  \label{f(T)}
\end{equation}
\begin{eqnarray}
F(T)=\frac{2\pi^2 \langle \delta q^2(T)\rangle }{e^2} &=&\frac{\pi}{e^2R_0}
\int\limits_{-\infty }^{+\infty }dx\frac{x\coth \left( \frac x{2TR_0C}
\right) }{1+x^2}=  \label{dq2(0)} \\
\  &=&F(0)+\frac{\pi ^2T}{E_C}+4\alpha _t\left( \ln \left( \frac{2\alpha
_tE_C}{\pi ^2T}\right) -\Psi \left( 1+\frac{2\alpha _tE_C}{\pi ^2T}\right)
\right) ,  \nonumber
\end{eqnarray}
\begin{equation}
g(T)=\frac{2e^2}\pi \int\limits_0^{+\infty }dt\left( \frac{\pi T}{\sinh \pi
Tt}\right) ^2t\left( K(t)\left( \cosh u(t)-1\right) +\frac{2\pi C}{e^2}\dot K
(t)\sinh u(t)\right) .  \label{g(T)}
\end{equation}
Here $\Psi (x)$ is the logarithm of the gamma-function, $\gamma =0.577...$
is the Euler constant and
\begin{eqnarray}
K(t) &=&R_0\theta (t)(1-\exp (-t/R_0C)),  \label{u(t)} \\
u(t) &=&R_0C\int d\omega \frac{\coth \frac \omega {2T}\sin \omega t}{
1+\omega ^2R_0^2C^2}.  \nonumber
\end{eqnarray}
Note that the value $F(0) \propto \langle \delta q^2(0)\rangle$ (\ref{dq2(0)})
diverges logarithmically at high
frequencies. As in the small voltage limit the Langevin equation approach
does not work at very
low $T$ (see below), in order to define $F(0)$ (or, equivalently, the
high frequency cutoff for (\ref{dq2(0)})) we should make use of a more
rigorous technique. For $V_x=0$ we find
\[
\left\langle \cos \frac{2\pi \hat q}e\right\rangle =\frac{\int d\varphi \rho
_{\text{eq}}(4\pi +\varphi ,\varphi )}{\int d\varphi \rho _{\text{eq}
}(\varphi ,\varphi )}\cos \frac{2\pi Q_g}e,
\]
$\rho _{\text{eq}}(\varphi
,\varphi ^{\prime })$ is the equilibrium density matrix of our system.
In the limit $\alpha_t >1$ this matrix  was evaluated by means of various
nonperturbative approaches \cite{PZ91,FSZ}.  In the leading order in
$e^{-2\alpha _t}$ all these approaches yield $\frac{\int d\varphi \rho _{\text{eq}
}(4\pi +\varphi ,\varphi )}{\int d\varphi \rho _{\text{eq}}(\varphi ,\varphi
)}=e^{-2\alpha _t}$. Thus we obtain
\begin{equation}
F(0)\simeq 2\alpha _t.  \label{cutoff}
\end{equation}
This equation completes our results. In the limit of large $T$ the
$Q_g$-dependent part of the conductance vanishes and we find the asymptotic behavior:
\begin{equation}
(R_R+R_L)G(T)=1-\frac{E_C}{3T}+\frac{6\zeta (3)}{\pi ^4}\alpha _t\left(
\frac{E_C}T\right) ^2-...  \label{asympt}
\end{equation}
At lower temperatures the conductance suppression due to charging effects
becomes more pronounced (fig. 1). Furthermore, by changing the gate charge
$Q_g$ it becomes possible to $e$-periodically tune the value of $G$. The
minimum and maximum conductance values (fig. 1) correspond to $Q_g=0$ and
$Q_g=e/2$ where the Coulomb barrier for electron tunneling reaches
respectively its maximum and minimum values. The modulation of $G$ with
$Q_g$ also becomes more pronounced as the temperature is lowered (fig. 2).

Finally, let us formulate the validity condition for our results. Analogously
to \cite{GZ92} we find that the phase fluctuations are sufficiently small
provided $T \gg E_C$ for $\alpha_t \ll 1$ and
\begin{equation}
T \gg \alpha_t E_C\exp (-2\alpha_t)
\label{val}
\end{equation}
for $\alpha_t \gtrsim 1$. Another our assumption $\langle \delta q^2\rangle \gtrsim e^2$
is also well justified for such values of $T$. According to (\ref{val}) the
validity domain of our analysis expands rapidly with increasing $\alpha_t$.
E.g. for the parameters of figs. 1,2 the condition (\ref{val}) yields
$T \gtrsim 20$ mK. And indeed our results show a very good agreement with
a preliminary experimental data of the Saclay group \cite{E} down to such
small values of $T$.

We would like to thank D.Esteve, H.Schoeller and G.Sch\"on for useful
discussions and comments. One of us (D.G.) acknowledges the support from
the International Center for Fundamental Physics in Moscow.

\begin{figure}[tbp]
\caption{Maximum and minimum conductance versus temperature ($E_C=0.715$K, $
\alpha_t=2.12$) }
\label{fig.1}
\caption{Dependence of conductance on gate voltage ($E_C=0.715$K, $
\alpha_t=2.12$)}
\label{fig.2}
\end{figure}

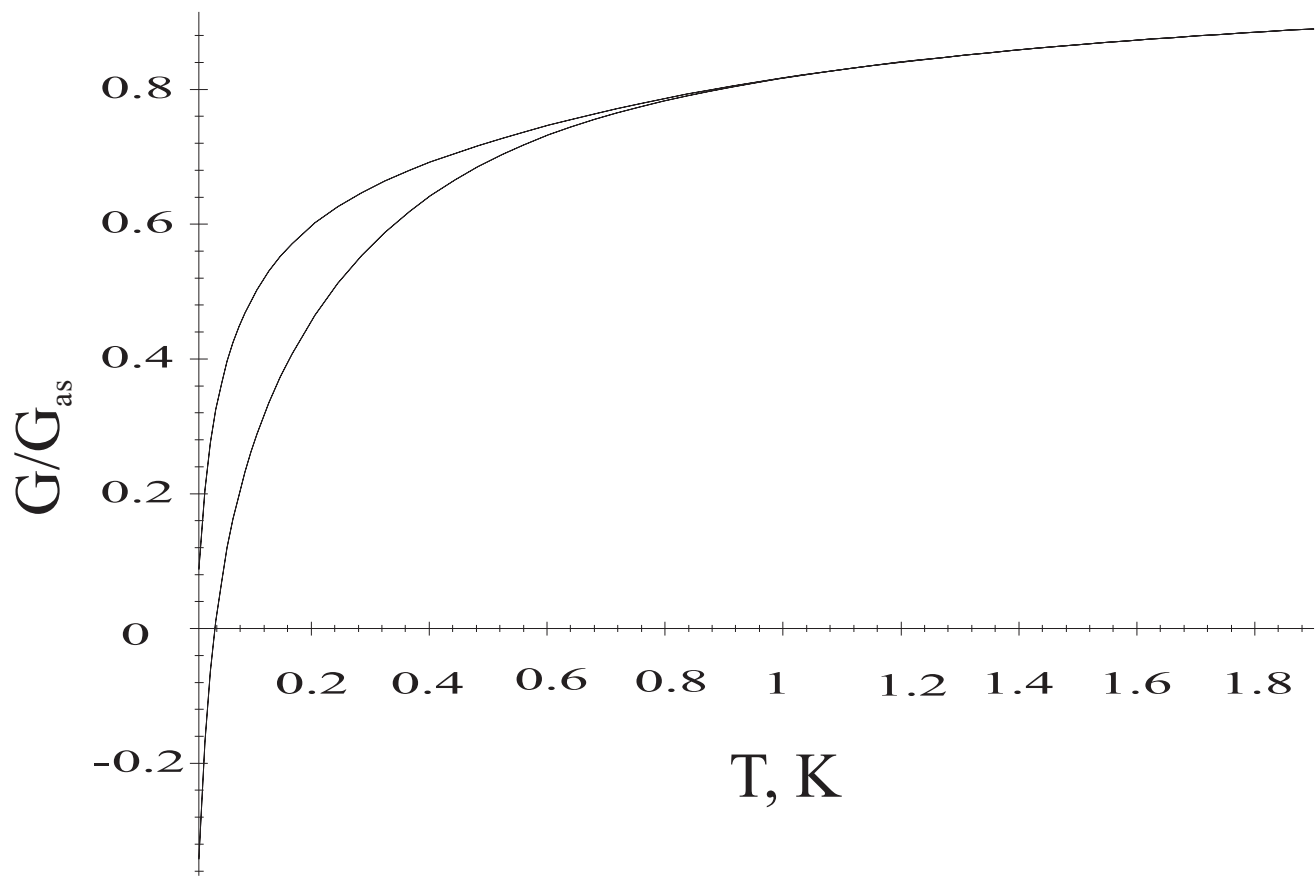
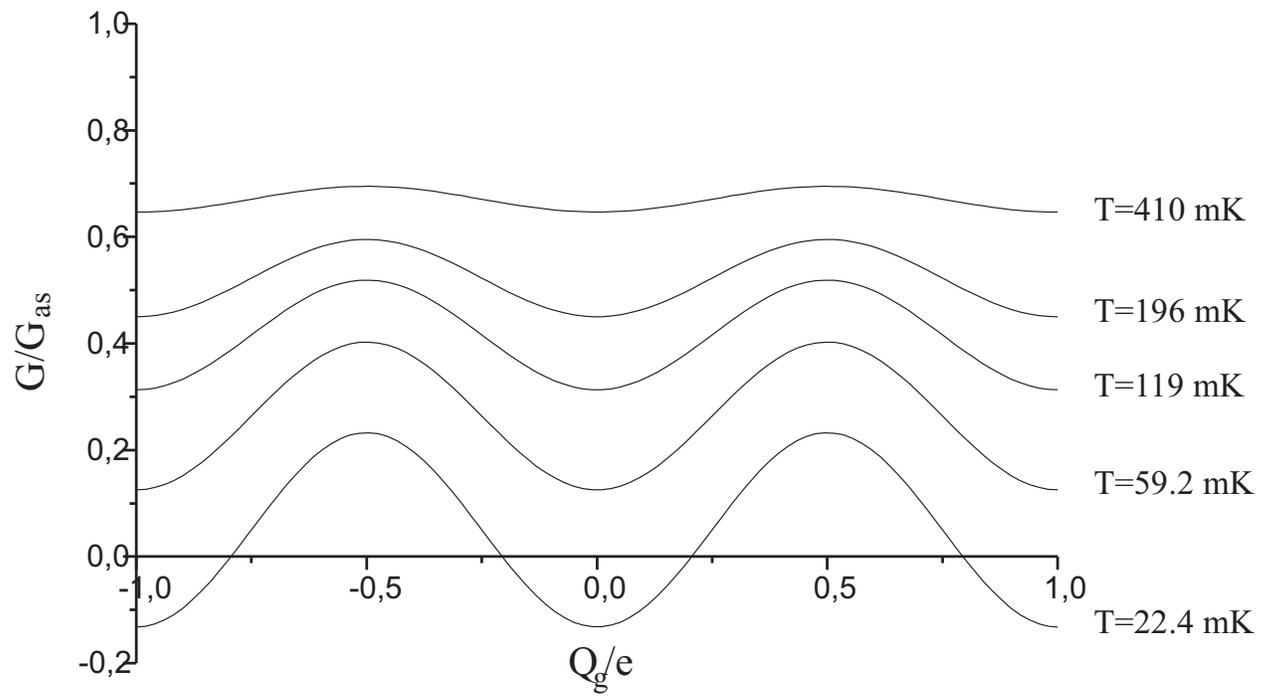
\end{document}